\documentclass[twocolumn,floatfix,superscriptaddress,aps,prb]{revtex4-1}
\usepackage{graphicx}
\usepackage{amsmath,mathtools}
\usepackage{amssymb}
\usepackage{bm}
\usepackage{hyperref}
\usepackage{color}

\newcommand{\n}[1]{\|#1 \|} 

\begin{document}

\title{Planckian superconductor}

\author{Y. Cheipesh}
\affiliation{Instituut-Lorentz, Universiteit Leiden, P.O. Box 9506, NL-2300 RA Leiden, The Netherlands}

\author{A. I. Pavlov}
\affiliation{The Abdus Salam International Centre for Theoretical Physics (ICTP) Strada Costiera 11, I-34151 Trieste, Italy}
\affiliation{Institute for Theoretical Solid State Physics, Leibniz-Institut f{\"u}r Festk{\"o}rper und
Werkstoffforschung IFW-Dresden, Helmholtzstrasse 20, D-01169 Dresden, Germany}

\author{V. Scopelliti}
\affiliation{Instituut-Lorentz, Universiteit Leiden, P.O. Box 9506, NL-2300 RA Leiden, The Netherlands}

\author{J. Tworzyd{\l}o}
\affiliation{Institute of Theoretical Physics, Faculty of Physics, University of Warsaw, ul.\ Pasteura 5, PL-02-093 Warszawa, Poland}

\author{N. V. Gnezdilov}
\email{nikolay.gnezdilov@yale.edu}
\affiliation{Department of Physics, Yale University, New Haven, CT 06520, USA}

\begin{abstract}
The Planckian relaxation rate $\hbar/t_\mathrm{P} = 2\pi  k_\mathrm{B} T$ sets a characteristic timescale for both the equilibration of quantum critical systems and maximal quantum chaos. In this note, we show that at the critical coupling between a superconducting dot and the complex Sachdev-Ye-Kitaev model, known to be maximally chaotic, the pairing gap $\Delta$ behaves as $\eta \, \hbar/t_\mathrm{P}$ at low temperatures, where $\eta$ is an order one constant.  The lower critical temperature emerges with a further increase of the coupling strength so that the finite $\Delta$ domain is settled between the two critical temperatures.
\end{abstract}

\date{\today}

\maketitle
 
The Bardeen-Cooper-Schrieffer mechanism of conventional superconductivity \cite{Bar57} requires two species of fermions coupled by an attractive two-body interaction. \cite{Alt10} The mean-field analysis of such a model results in the gapped quasiparticle excitation spectrum below the critical temperature. Meanwhile, the absence of long-living quasiparticles in high-temperature superconducting materials \emph{above} the critical temperature is an immutable characteristic of the so-called strange metal state. \cite{Sen08,Kei15} In contrast to the quasiparticle nature of superconductors, strange metals exhibit a power-law behavior in the spectral function, \cite{Var89} similarly to quantum critical systems. \cite{Sac11}
A lack of quasiparticles manifests itself in fast equilibration at low temperature on a timescale set by the Planckian relaxation time $t_\mathrm{P} = \hbar/\left(2\pi  k_\mathrm{B} T\right)$. \cite{Zaa04,Sac11} The same timescale appears as an upper bound on quantum chaos setting the maximal rate of information scrambling. \cite{Mal16_2} It is usually formulated \cite{Mal16_2,She15,Rob16} in terms of the out-of-time ordered correlator \cite{Lar69} (OTOC): In quantum many-body systems the OTOC grows no faster than exponentially $e^{t/t_\mathrm{L}}$ with the Lyapunov time $t_\mathrm{L}$ bounded from below as $t_\mathrm{L} \geq t_\mathrm{P}$. \cite{Mal16_2} 

The widely known Sachdev-Ye-Kitaev (SYK) model, \cite{Sac93,Kit15} describing strongly interacting Majorana zero modes in $0+1$ dimensions, saturates the chaos bound $t_\mathrm{L}=t_\mathrm{P}$. \cite{Kit15,Mal16_1}  It does not possess an underlying quasiparticle description while being solvable in the infrared, with a spectral function that scales as a power law of frequency. 
These properties do not change upon replacing Majoranas with conventional fermions (complex SYK model). \cite{Sac15,Bul17} The extensions of this model to the cSYK coupled clusters predict thermal diffusivity \cite{Dav17} $\propto t_\mathrm{P}$ and  reproduce the linear in temperature resistivity, \cite{Son17} observed in strange metals. \cite{Tak92, Tai10}
Recently, a proposed theory of a Planckian metal, \cite{Pat19} based on the destruction of a Fermi surface by the cSYK-like interactions, shows that the universal scattering time equals the Planckian time $t_\mathrm{P}$. The latter one characterizes the linear in temperature resistivity property \cite{Bru13} and was detected in cuprates, \cite{Leg19} pnictides, \cite{Nak19} and twisted bilayer graphene, \cite{Cao19} regardless of their different microscopic nature.

The success in applying the SYK model to qualitative studies of strange metals and the minimalistic structure of the model itself fostered the effort to find a mechanism by which the superconducting state is formed out of an incoherent SYK metal. \cite{Pat18sc, Est19, Wan19, Deb19}
Driven by the same curiosity, we consider a $(0+1)$-dimensional toy model which consists of a superconducting quantum dot \cite{Kou98} coupled to the complex-valued SYK model. \cite{Sac15}
At the critical coupling the pairing gap turns out to be proportional to the Planckian relaxation rate at low temperatures,
\begin{equation}
\Delta \approx \eta \, \frac{\hbar}{t_\mathrm{P}}, \label{Delta_c}
\end{equation}
where $\eta$ is a number close to one. This theoretical finding that we refer to as a Planckian superconductor draws parallels to the phenomenon of reentrant superconductivity \cite{Map72,Sim12} in Kondo superconductors \cite{Mul71,Rib71,Mul76} and the physics of Andreev billiards. \cite{Mel97,Sch99,Lod98,Ada02,Vav03}

We start with a superconducting Hamiltonian $H_{\rm SC}$ that contains $2M$ modes described by the Richardson Hamiltonian \cite{Ric63, Del96, Mat97} without single-particle energies coupled to the SYK model $H_{\rm SYK}$ with $N$ fermions through a random tunneling term $H_{\rm tun}$,
\begin{align}
H\!=&{}H_{\rm SC} + H_{\rm SYK}+ H_{\rm tun}\, \label{H}, \\
H_{\rm SC}\!=&{}-\frac{U}{M} \!\sum_{i,j=1}^M \!\!\psi^\dag_{\uparrow i} \psi^\dag_{\downarrow i} \psi_{\downarrow j} \psi_{\uparrow j}-\mu\! \sum_{i=1}^M \!\sum_{\sigma=\uparrow,\downarrow} \!\!\psi^\dag_{\sigma i} \psi_{\sigma i}  \label{H_BCS},\\
H_{\rm SYK}\!=&{}\frac{1}{(2N)^{3/2}}\!\sum_{i,j,k,l=1}^N \!\!J_{ij;kl} c^\dag_i c^\dag_j c_k c_l \label{H_SYK}, \\
H_{\rm tun}\!=&{}\frac{1}{(MN)^{1/4}} \!\sum_{i=1}^N \sum_{j=1}^M \!\sum_{\sigma=\uparrow,\downarrow}\!\!\left(t^\sigma_{ij} c^\dag_i\psi_{\sigma j}  +\mathrm{h}.\mathrm{c}. \right) \label{H_tun}.
\end{align} 
The couplings $t^\sigma_{ij}$ and $J_{ij;kl}$ are assumed to be independent Gaussian random variables with finite variances $\overline{{t^\sigma}^*_{ij}t^{\sigma'}_{ij}}=t^2 \delta_{\sigma\sigma'}$, $\overline{|J_{ij;kl}|^2}=J^2$ ($J_{ij;kl}=-J_{ji;kl}=-J_{ij;lk}=J^*_{kl;ij}$), and zero means. 

The interaction terms in the Hamiltonian (\ref{H}) are decoupled within the Hubbard-Stratonovich transformations, \cite{Alt10,Sac15}  so that
in the large $M,N$ limit the self-consistent saddle-point equations are \cite{appA}
\begin{align}
\Sigma_c(\tau)&=J^2 G_c(\tau)^3 +2 \sqrt{p} \, t^2 G_+(\tau)\label{Sc}\,,\\
G_c(\mathrm{i}\omega_n)^{-1}&=\mathrm{i}\omega_n -\Sigma_c(\mathrm{i}\omega_n)\label{Gc},\\
G_+(\mathrm{i}\omega_n)&=\frac{\mathrm{i}\omega_n -\frac{t^2}{\sqrt{p}} G_c(\mathrm{i}\omega_n)}{\left(\mathrm{i}\omega_n-\frac{t^2}{\sqrt{p}} G_c(\mathrm{i}\omega_n) \right)^{\!2}\!\!-|\Delta|^2}\label{G+},\\
\frac{1}{U}&= T\!\!\sum_{n=-\infty}^{+\infty} \! \frac{1}{\left( \omega_n +\frac{\mathrm{i} t^2}{\sqrt{p}} G_c(\mathrm{i}\omega_n)\!\right)^{\!2}\ \!\!+|\Delta|^2}\label{delta},
\end{align} 
where $\omega_n =\pi T(2n+1)$ are Matsubara frequencies and $p=M/N$ controls the ratio between the ``sites'' \cite{Ban17, Che17, Jia17} in the superconductor/SYK sector.
The self-energy of the SYK fermions appears in the equations (\ref{Sc},\ref{Gc}) as $\Sigma_c(\tau)$, while $G_c(\tau)$ denotes the corresponding Green's function $-N^{-1} \sum_{i=1}^N \left\langle \mathrm{T}_\tau c_i(\tau) \bar{c}_i(0)\right\rangle$.
The Green's functions of the $\uparrow$,$\downarrow$ fermions in the superconductor $G_\sigma(\tau)=-M^{-1} \sum_{i=1}^M \left\langle \mathrm{T}_\tau \psi_{i\sigma}(\tau) \bar{\psi}_{i\sigma}(0)\right\rangle$ enter the equation (\ref{G+}) as a half trace of the Gor'kov's function \cite{Gor58}  $G_+(\tau) = \tfrac{1}{2}(G_\uparrow+G_\downarrow)(\tau)$.
Finally, relation (\ref{delta}) is a modified gap equation, \cite{Alt10} which accounts for the amount of the SYK impurity in the superconductor through $G_c(\tau)$ under the assumption of frequency independent pairing $\Delta$.  
The chemical potential $\mu$ can be accounted in the equations (\ref{Sc}-\ref{delta}) by the shift $|\Delta|^2 \to |\Delta|^2+\mu^2$.
Below, we set $\mu=0$. 

\begin{figure}[t!!]
\center
\includegraphics[width=0.976\linewidth]{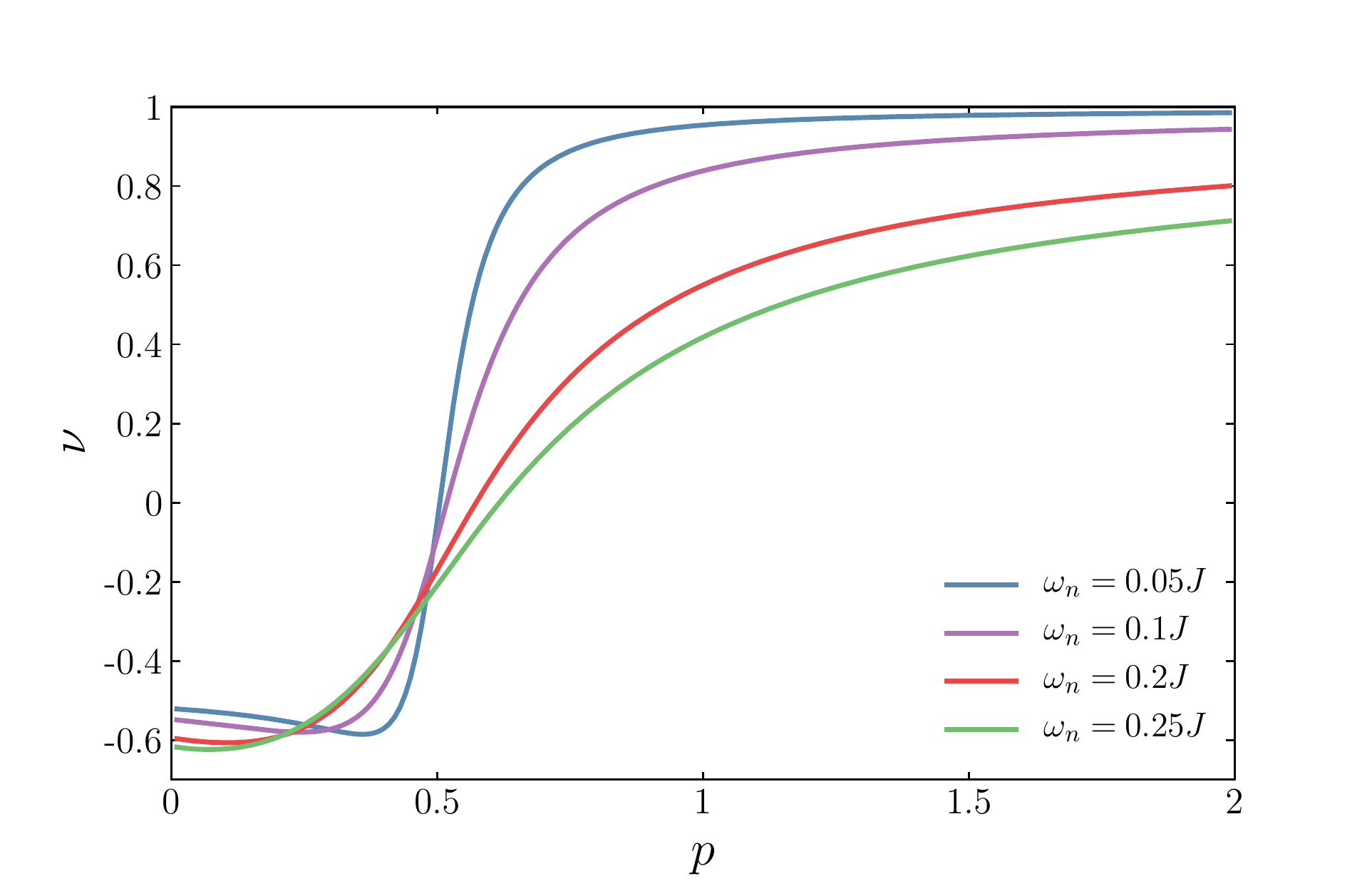}
\caption{\small \label{fig:nu} \textbf{Scaling of the Green's function} $G_c$ in the normal phase. We plot $\nu = \partial \ln G_c/\partial \ln \omega_n$ as a function of $p$ at given frequencies and finite coupling $t=0.475 J$. At low frequencies, $\nu$ close to $-1/2$ is robust against $p$ increase for $p<1/2$. The frequency rise moves $\nu$ towards $-1$ (free fermion limit), while $\nu$ crosses over to $1$ for large $p$.  The temperature is $T=10^{-4} J$.}
\end{figure}

In the normal phase ($\Delta=0$) the equations (\ref{Sc}-\ref{G+}) can be written as
\begin{align}
\Sigma(\tau)&=J^2 G_c(\tau)^3\label{Sc_shifted},\\
\left(\mathrm{i}\omega_n -\Sigma(\mathrm{i}\omega_n)\right)G_c(\mathrm{i}\omega_n)&=\frac{\mathrm{i}\omega_n -\frac{t^2\left(1-2p\right)}{\sqrt{p}}G_c(\mathrm{i}\omega_n)}{\mathrm{i}\omega_n-\frac{t^2}{\sqrt{p}} G_c(\mathrm{i}\omega_n) }\label{Gc_eq},
\end{align} 
ensuring a convenient self-energy translation $\Sigma\equiv \Sigma_c-2 \sqrt{p} \, t^2 G_+$. 
If $p \ll 1/2$ ($2M \ll N$), the bare SYK Green's function $G_{\rm SYK}(\mathrm{i}\omega_n)=-\mathrm{i}\pi^{1/4}\mathrm{sgn}\left(\omega_n\right)/\sqrt{|J \omega_n|}$ solves the equations (\ref{Sc_shifted},\ref{Gc_eq}) in the infrared $\omega_n \ll J$. In this regime, the Green's function of the $\psi$ fermions $G_+(\mathrm{i}\omega_n)$ scales as $\sqrt{\omega_n}$ for $\omega_n/J \ll p^{-1/3}(t/J)^{4/3}$.
In the equal sites case $2M=N$, which corresponds to $p=1/2$, the bare SYK Green's function survives for $\left(t/J\right)^{4/3}\ll \omega_n/J \ll 1$.  Another solution appears at $p=1/2$ if one supposes $\omega_n \ll \left\lbrace t^2 \left|G_c\right|, \left|\Sigma\right| \right\rbrace$. Then the equation (\ref{Gc_eq}) shortens to 
\begin{align}
\Sigma(\mathrm{i}\omega_n)&=\frac{\mathrm{i}\omega_n}{\sqrt{2}t^2}\,   G_c(\mathrm{i}\omega_n)^{\!-2} \label{Gc_eq_05}.
\end{align} 
The Green's function that satisfies the equations (\ref{Sc_shifted},\ref{Gc_eq_05}) is $G_c(\mathrm{i}\omega)\propto -\mathrm{i}\,\mathrm{sgn}(\omega)/\left(J^2 t^2 |\omega_n|\right)^{\!-1/5}$ for the frequencies $\left(t/J\right)^3\ll \omega_n/J \ll \left(t/J\right)^{4/3}$, that are achievable in the weak tunneling limit $t \ll J$. Note that the frequency window strictly depends on the coupling $t$. 
For $p \gg 1/2$, the Green's function of the $c$ fermions in the low-frequency limit is $G_c(\mathrm{i}\omega_n)\propto -\mathrm{i} \omega_n$, \cite{Jia17} which leads to the density of states $-\pi^{-1}\mathrm{Im} G_c(\mathrm{i}\omega_n \to \omega+\mathrm{i}0^+)\simeq 0$ vanishing in the SYK sector. 
Therefore, at large $p$, the normal phase is given by the free fermions in the $\psi$--dot, whose Green's function is $G_+(\mathrm{i}\omega_n)=-\mathrm{i}/\omega_n$.  To follow the frequency scaling of the Green's function $G_c(\mathrm{i}\omega_n)$ while changing $p$, we introduce the logarithmic derivative $\nu=\partial \ln G_c/\partial \ln \omega_n$ plotted in Figure \ref{fig:nu} at low temperatures.
Summarizing, the normal phase in the infrared limit is described by the inverse Green's function of the SYK model at small $p$, whereas it crosses over to free fermions for large $p$ values. 

\begin{figure*}[t!!]
\center
\includegraphics[width=0.488\linewidth]{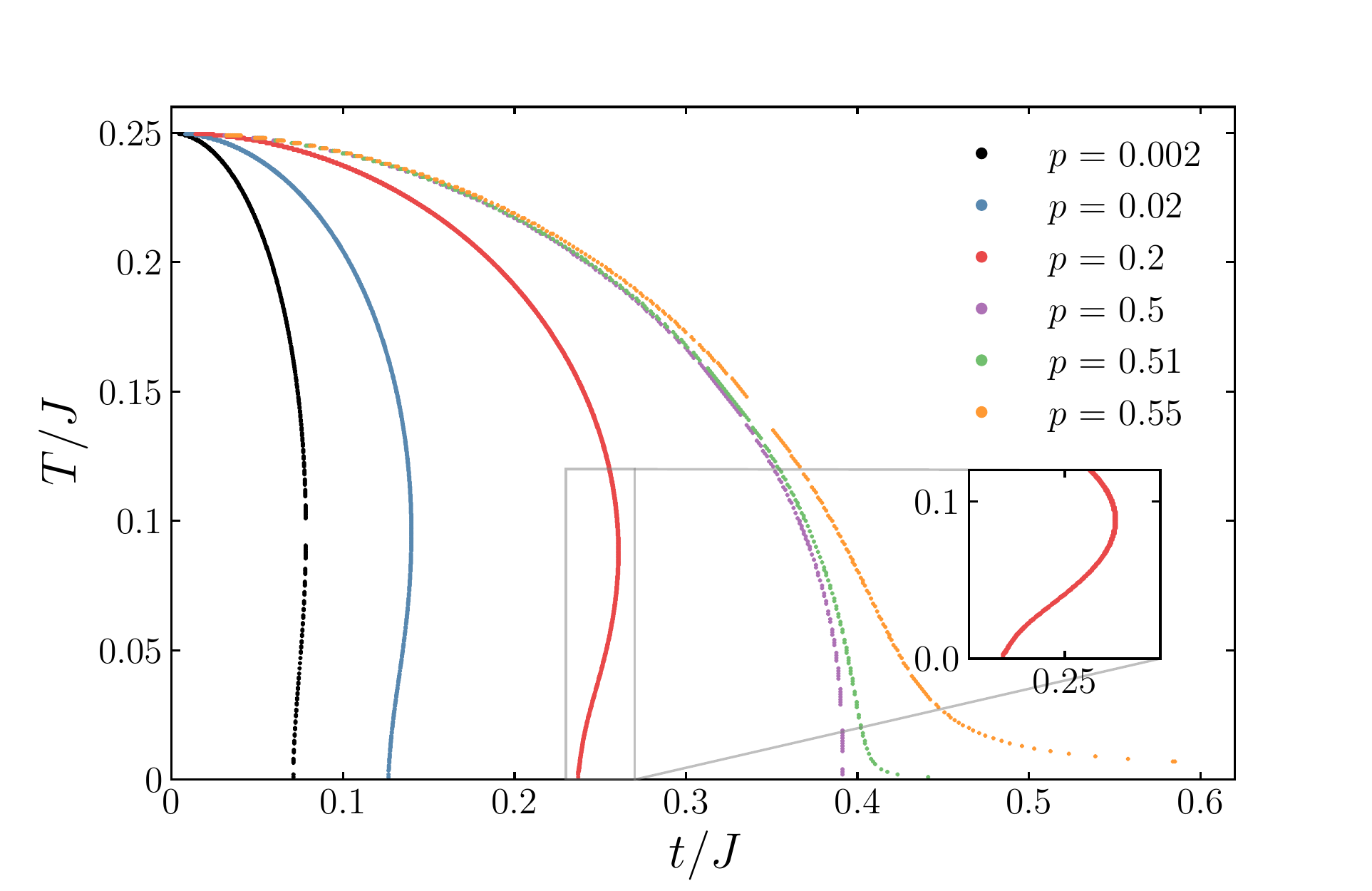} \quad
\includegraphics[width=0.488\linewidth]{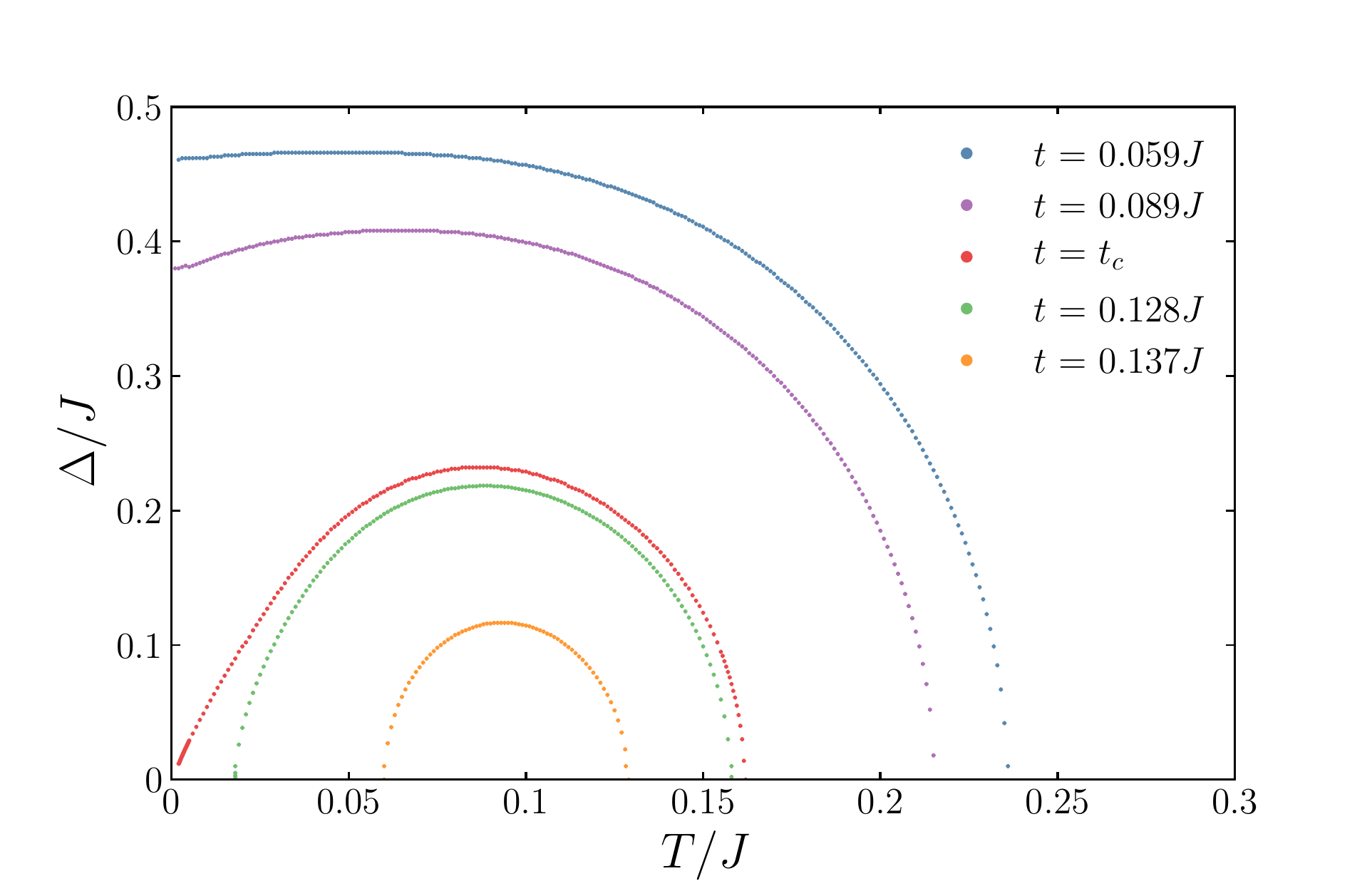}
\caption{\small \label{fig:Tt_delta} \textbf{Left panel: Critical temperature} as a function of the coupling strength to the SYK dot. The curves for $p<0.5$ are bent at low temperatures. This illustrates the presence of \textbf{two critical temperatures}. At $p=0.5$ the bend disappears, whereas for the values of $p>0.5$ a single critical temperature decays to zero asymptotically.
\textbf{Right panel: The pairing gap} as a function of temperature at $p=0.02$. The critical coupling value is $t_c \approx0.127 J$.
$U$ is set equal to $J$ in both panels. }
\end{figure*}

The gap equation (\ref{delta}) at $\Delta=0$ makes a boundary in between the normal phase and the superconducting one by setting the critical temperature $T_c$ as a function of the coupling rate $t$. 
Let us notice that the SYK model (\ref{H_SYK}) does not have a spin degree of freedom after disorder averaging. \cite{appA} Thus, it may be thought of as spin polarized. It suppresses superconductivity similar to magnetic impurities: Increase of the coupling to the SYK subsystem decreases the critical temperature. \cite{deg66} There exists a critical coupling $t_c$, 
\begin{equation}
\frac{1}{U}=\!\int_{-\infty}^{+\infty} \frac{d\omega}{2\pi} \! \left( \omega +\frac{\mathrm{i} t_c^2}{\sqrt{p}} G_c(\omega)\!\right)^{\!-2} \label{tc},
\end{equation} 
such as to abolish superconductivity at zero temperature. The constraint (\ref{tc}) follows from the gap equation (\ref{delta}) when $\Delta,T=0$. 

\begin{figure}[b!!]
\center
\includegraphics[width=0.976\linewidth]{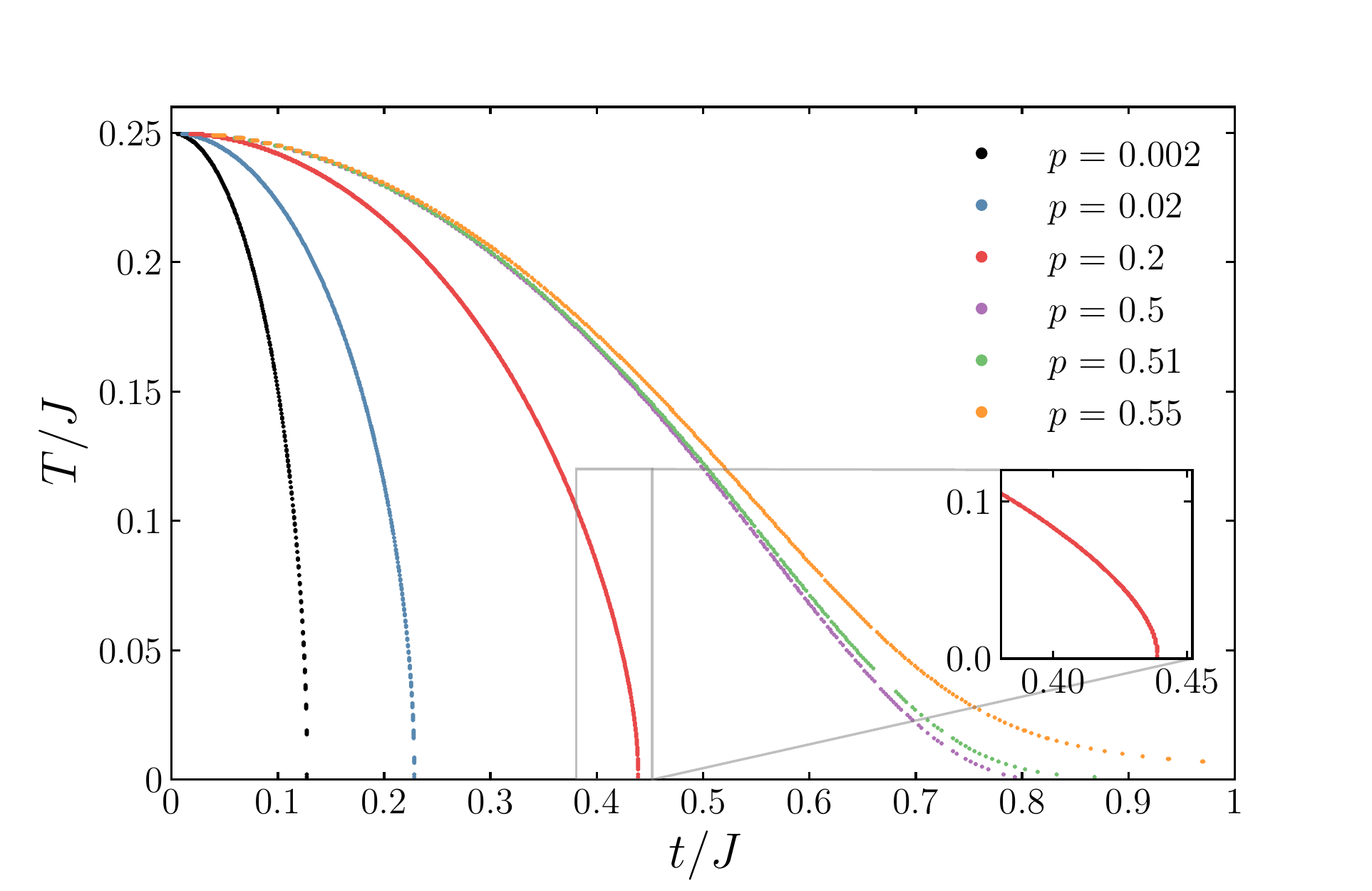}
\caption{\small \label{fig:Tt_dFl} \textbf{Critical temperature} as a function of the coupling strength to the \textbf{random free fermions} model.}
\end{figure}

There are three competing phases contributing to the denominator of the self-consistency relation (\ref{delta}): SYK non-Fermi liquid, free fermions, and superconducting condensate $\Delta$. If there are enough of the SYK fermions ($N > 2M$), $\Delta$ interplays with the non-Fermi liquid at zero temperature. The latter one falls off with an increase in temperature, making room for the superconducting phase beyond the critical coupling, which results in the growth of the critical temperature. Indeed, Figure \ref{fig:Tt_delta} (left) shows the bend of the critical temperature in the vicinity of the critical coupling. \cite{comment}  This phenomenon resembles the reentrant superconductivity \cite{Map72,Sim12} in superconductors with Kondo impurities. \cite{Mul71,Rib71,Mul76} The pairing gap goes down at low temperatures with an increase in coupling as in Figure \ref{fig:Tt_delta} (right). Achieving the critical coupling when $\Delta$ vanishes at zero temperature leads to the appearance of the lower critical temperature. 
In contrast, the reentrant superconducting regime is absent for $N < 2 M$, since the normal phase behaves as the conventional Fermi liquid at low temperatures and large $p$, as was noticed earlier. In Figure \ref{fig:Tt_delta} (left), we show \cite{comment} that $p=1/2$ ($N=2 M$) separates the regions with one or two critical temperatures. Similarly, consideration of the random free fermion model $\sum_{ij} J_{ij} c^\dag_i c_j$ instead of the SYK model does not give the reentrance effect. In this case, the self-energy equation (\ref{Sc}) changes to $\Sigma_c(\mathrm{i} \omega_n)=J^2 G_c(\mathrm{i} \omega_n) +2 \sqrt{p} \, t^2  G_+(\mathrm{i} \omega_n)$. The results for the critical temperature are presented in Figure \ref{fig:Tt_dFl}. It is still possible to suppress the superconductivity at zero temperature providing sufficient impurities, but there is only a single critical temperature as the normal phase is always set by the free fermions. \cite{comment_unparticle}

\begin{figure*}[t!!]
\center
\includegraphics[width=0.488\linewidth]{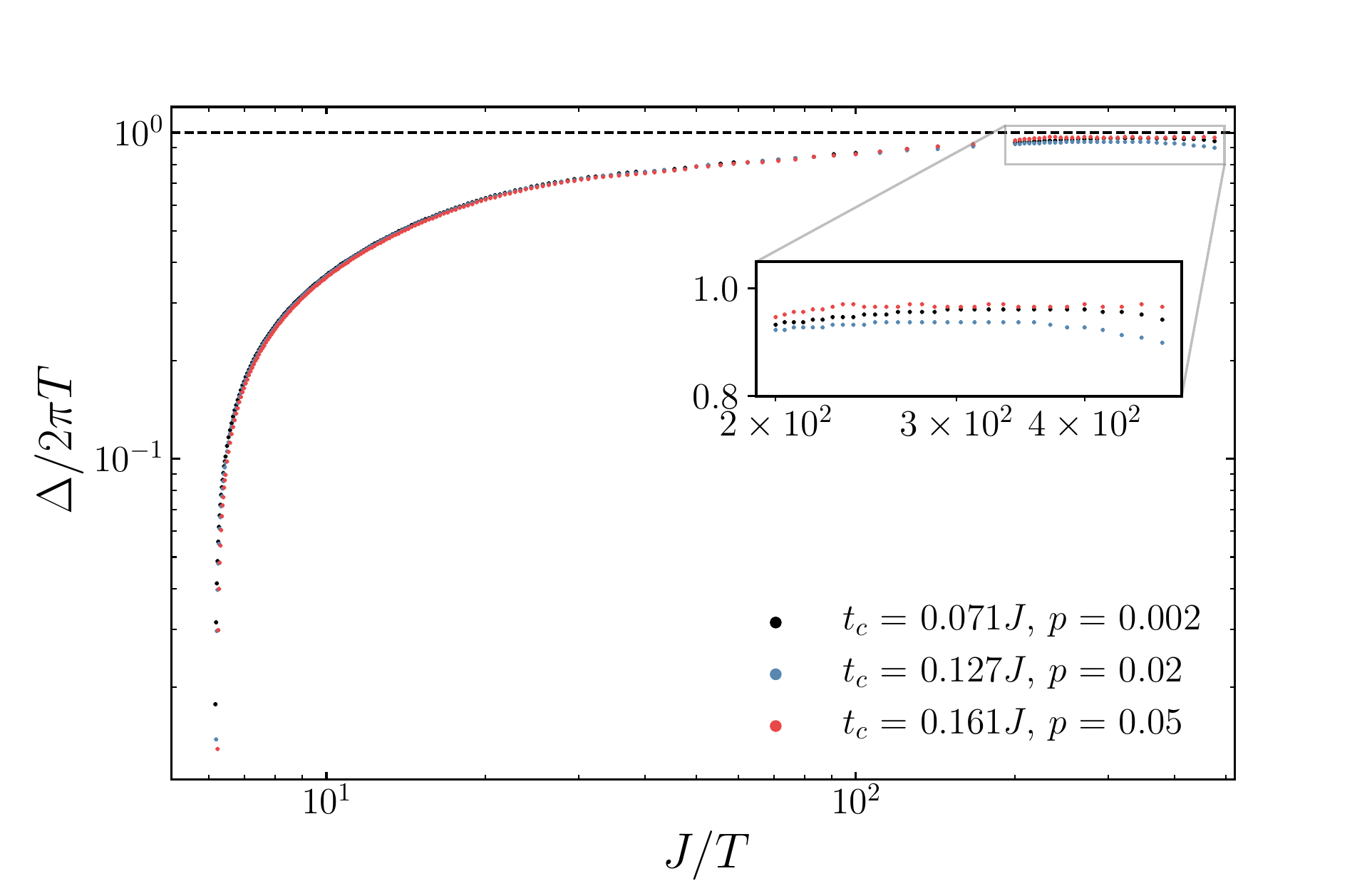} \quad
\includegraphics[width=0.488\linewidth]{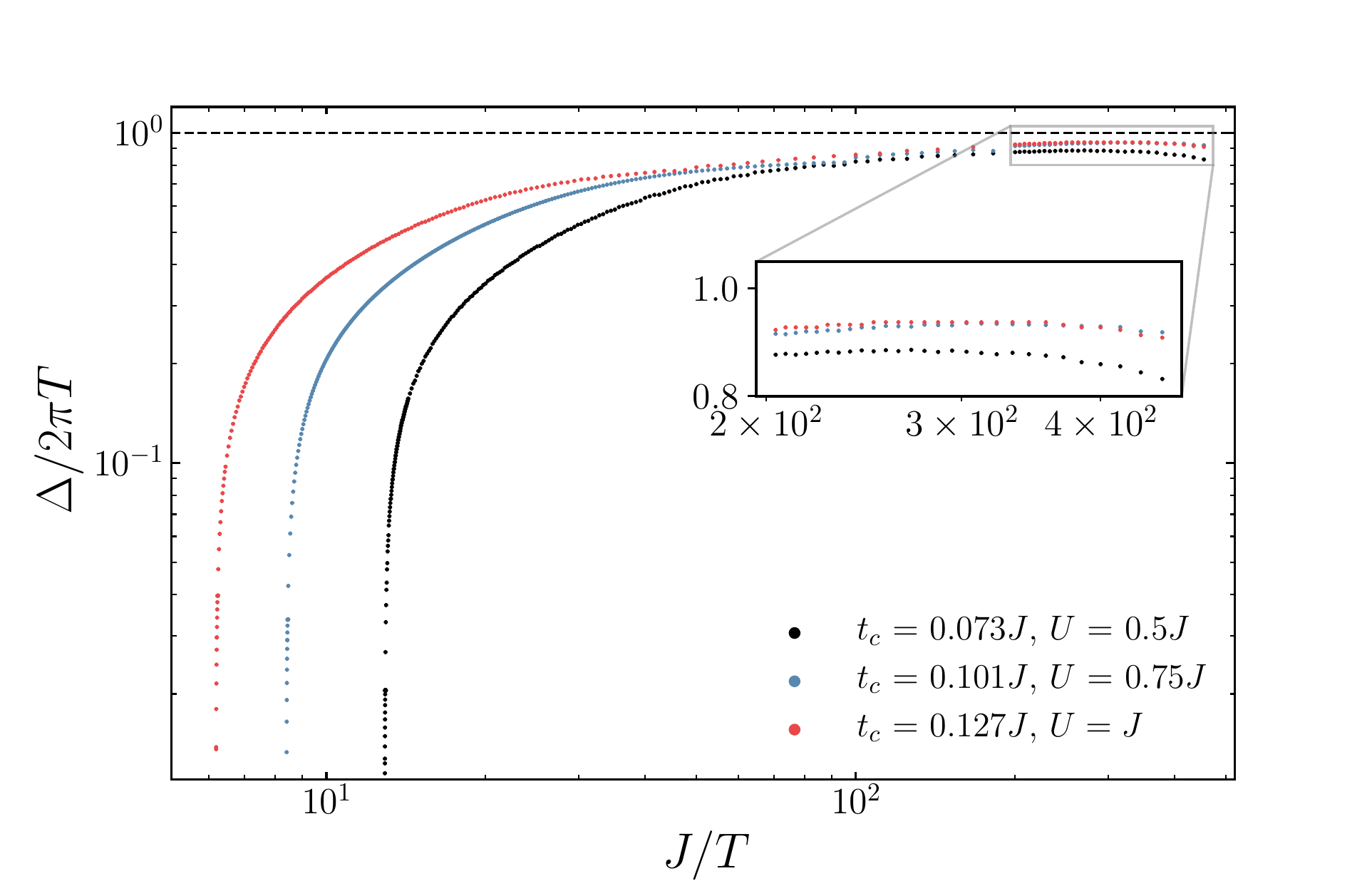}
\caption{\small \label{fig:delta/T} \textbf{The gap to temperature ratio} as a function of inverse temperature at the critical coupling depends on neither the mode ratio $p$ (fixed $U=J$, \textbf{left panel}) nor the Richardson interaction strength $U$ (fixed $p=0.02$, \textbf{right panel}). In both cases, $\Delta$ saturates $2 \pi T$ at low temperatures. \cite{comment_fig} In the right panel, we notice that a decrease of the interaction in the superconducting dot reduces the critical temperature as in the bare Richardson model (\ref{H_BCS}).}
\end{figure*}

From Figure \ref{fig:Tt_delta} (right), one notices the pairing gap at the critical coupling is $\propto T$ at low temperatures. We numerically examine \cite{comment} $\Delta$ in the reentrant phase $p< 1/2$ for several values of $p$ and $U$ (see Figure \ref{fig:delta/T}). The gap saturates $2 \pi T$ almost irrespective of parameters of the problem. Unit recovery brings us to the above-mentioned relation (\ref{Delta_c}) so that the gap is set by the inverse Planckian time $1/t_\mathrm{P}$ multiplied by $\hbar$. 

This observation seems to be reminiscent of quite a peculiar feature of an Andreev billiard: \cite{Bee05} In a clean chaotic cavity proximate to a superconductor, the induced gap equals $\hbar/t_\mathrm{E}=\hbar/\!\left(t_\mathrm{L}\ln \frac{p_\mathrm{F}  l}{\hbar}\right)$,\cite{Lod98,Ada02,Vav03} where $t_\mathrm{E}$ is the Ehrenfest time (the typical timescale of quantum dynamics), $t_\mathrm{L}$ is the Lyapunov time, $p_\mathrm{F}$ is the Fermi momentum, and $l$ is the characteristic cavity length. The effect is predicted in the regime of the Ehrenfest time far exceeds $\tau$ the typical lifetime of an electron/hole excitation in the cavity. 
Oppositely, if $t_\mathrm{E} \ll \tau$, the gap behaves as $\hbar/\tau$, where $\tau$ does not depend on the Planck constant. \cite{Mel97,Sch99}
In the SYK model the Lyapunov time coincides with the Planckian relaxation time  $t_\mathrm{L}=\hbar/\left(2 \pi k_\mathrm{B}T \right)=t_\mathrm{P}$, \cite{Kit15,Mal16_1} although those are different physical quantities. \cite{comment_time} However, the Ehrenfest time is $t_\mathrm{L} \ln N \gg \hbar/(2 \pi k_\mathrm{B} T)$, which differs from $t_\mathrm{P}$ predicted in the pairing gap (\ref{Delta_c}) by $\ln N$. 

To estimate the gap behavior at the critical coupling we consider the equations (\ref{Sc}-\ref{G+}) at finite $\Delta$,
\begin{widetext}
\begin{align} 
&\Big(\mathrm{i}\omega_n -\Sigma(\mathrm{i}\omega_n)\Big)G_c(\mathrm{i}\omega_n)=\frac{\left(\mathrm{i}\omega_n\!-\!\frac{t^2}{\sqrt{p}} G_c(\mathrm{i}\omega_n)\!\right)\!\!\left(\mathrm{i}\omega_n\! -\!\frac{t^2\left(1-2p\right)}{\sqrt{p}}G_c(\mathrm{i}\omega_n)\!\right)\!-\!|\Delta|^2}{\left(\mathrm{i}\omega_n\!-\!\frac{t^2}{\sqrt{p}} G_c(\mathrm{i}\omega_n)\!\right)^{\!2}\!\!-\!|\Delta|^2}\label{Gc_eq_delta},
\end{align} 
\end{widetext}
whereas the self-energy equation (\ref{Sc_shifted}) stays unchanged.  The right-hand side of the equation (\ref{Gc_eq_delta}) tends to unity for $p\ll 1/2$. Thus it is sufficient to substitute the SYK Green's function in the gap equation (\ref{delta}) in this regime. 

As we look for a low-temperature correction to zero $\Delta$ at the critical coupling, we expand the gap equation (\ref{delta}) in powers of $\Delta$ up to the second order,
\begin{align}
\frac{1}{U} \!\simeq &\, 2T\!\sum_{n=0}^{+\infty} \! \frac{1}{\left( \!\omega_n\! +\!\frac{\mathrm{i} t_c^2}{\sqrt{p}} G_c(\omega_n)\!\right)^{\!\!2}} \!\!\left(\!\!1\!-\!\frac{|\Delta|^2}{\left( \!\omega_n\! +\!\frac{\mathrm{i} t_c^2}{\sqrt{p}} G_c(\omega_n)\!\right)^{\!\!2}}\!\!\right)\!\! \label{delta_2o}.
\end{align}
The SYK Green's function diverges at low frequencies as $1/\sqrt{\omega_n}$ and decays as $1/\omega_n$ in the ultraviolet. Hence the principal contribution to the sum (\ref{delta_2o}) from the high frequencies is given by the bare $\omega_n$ in the denominator. On the other hand,  a divergent Green's function is crucial at low frequencies. Assuming $G_c$ decays fast enough in comparison to $\omega_n$, we replace $G_c$ with the infrared SYK Green's function $G_{SY\!K}(\mathrm{i}\omega_n)=-\mathrm{i}\pi^{1/4}\mathrm{sgn}\left(\omega_n\right)/\sqrt{|J \omega_n|}$ in  expression (\ref{delta_2o}). 

The low-temperature version of relation (\ref{delta_2o}) can be written by means of the Euler-Maclaurin formula, \cite{Abr64}
\begin{align} \nonumber
\frac{1}{U} \simeq & \!\int_{0}^{+\infty}\!\!\frac{d\omega}{\pi} \! \frac{1}{\left( \!\omega\! +\!\frac{\mathrm{i} t_c^2}{\sqrt{p}} G_{SY\!K}(\omega)\!\right)^{\!\!2}} \!\!\left(\!\!1\!-\!\frac{|\Delta|^2}{\left( \!\omega\! +\!\frac{\mathrm{i} t_c^2}{\sqrt{p}} G_{SY\!K}(\omega)\!\right)^{\!\!2}}\!\!\right)\!\\&-\!\frac{pT}{ t_c^4 \, G_{SY\!K}(\pi T)^{2}}\!\left(\!1+\!\frac{2\pi T}{3}\frac{\partial G_{SY\!K}(\pi T)/\partial \omega}{G_{SY\!K}(\pi T)} \!\right) \!\label{EM},
\end{align}
where we expand up to $T^2$ keeping in mind that $\Delta \propto T$ at the critical coupling.  \cite{comment_T}
Finally, one notices two terms in the top row of the equation (\ref{EM}) that match the critical coupling condition (\ref{tc}). 
Therefore, we obtain \cite{comment_int}
\begin{align} 
\Delta(T)\simeq \, \sqrt{6}\pi T \label{delta_est}.
\end{align}
Although this estimate gives $\eta\approx 1.22$ that exceeds the found numerical value $\eta \approx 0.96$ for the pairing gap $\Delta\approx \eta \, \hbar/t_\mathrm{P}$,  the derived low-temperature gap behavior (\ref{delta_est}) is independent of the problem parameters as in Figure \ref{fig:delta/T}.

\textit{Conclusion}.---\,In this manuscript, we considered the superconducting proximity effect for the Sachdev-Ye-Kitaev model. We have shown, that the superconducting dot coupled to the complex SYK model possesses reentrant superconductivity. At the critical coupling, which gives rise to the occurrence of a lower critical temperature, the pairing gap disappears at $T=0$ and grows linearly with an increase in temperature. The linear--$T$ growth of the gap is given by  $\hbar/t_\mathrm{P}$, where $t_\mathrm{P}=\hbar/\left(2\pi k_\mathrm{B} T\right)$ is the Planckian relaxation time. The same timescale serves as an ultimate bound on many-body quantum chaos, \cite{Mal16_2} saturated in strongly coupled systems without quasiparticle excitations. Thereby a natural question arises whether the pairing gap is an appropriate physical observable for the Lyapunov spectrum \cite{Rom19} of the SYK model. Accurate studies of the OTOC in the proposed system (\ref{H}) might shed light on that. 
On its own, $\Delta \approx \eta \, \hbar/t_\mathrm{P}$ may be used to characterize the cSYK quantum dots. \cite{Pik18,Dan17}
However, this requires consideration of a more realistic setup such as a superconducting lead attached to the particular realization of the complex SYK model.    

\acknowledgments 
We are grateful to C. W. J. Beenakker for drawing our attention to this problem. 
The authors have benefited from inspiring discussions with D. V. Efremov, Yu. Malitsky, and K. E. Schalm. 
This research was supported by  the Netherlands Organization for Scientific
Research (NWO/OCW), by the European Research Council, and by the DOE Contract DEFG02-08ER46482 (NVG).

\begin{widetext}

\appendix

\section{Derivation of the gap equation}\label{appA}

The imaginary time action averaged over disorder is 
\begin{align} \nonumber
S=&{}\int_0^\beta \!\! d \tau \Bigg[\sum_{i=1}^N \bar{c}_i \partial_\tau c_i+\!\sum_{i=1}^M\!\sum_{\sigma=\uparrow,\downarrow}\!\!\bar{\psi}_{\sigma i} \left(\partial_\tau-\mu\right) \psi_{\sigma i}  -\frac{U}{M} \!\sum_{i,j=1}^M \!\!\bar{\psi}_{\uparrow i} \bar{\psi}_{\downarrow i} \psi_{\downarrow j} \psi_{\uparrow j} \Bigg] \\  &{}-\int_0^\beta \!\! d \tau \!\! \int_0^\beta \!\! d \tau' \Bigg[ \frac{t^2}{\sqrt{NM}}  \!\sum_{i=1}^N \sum_{j=1}^M \!\sum_{\sigma=\uparrow,\downarrow}\!\!\bar{c}_i \psi_{\sigma j} (\tau) \bar{\psi}_{\sigma j} c_i (\tau') 
- \frac{J^2}{4 N^3}\!\sum_{i,j,k,l=1}^N \!\bar{c}_i \bar{c}_j c_k c_l(\tau) \bar{c}_l \bar{c}_k c_j c_i(\tau')\Bigg] \label{appA:S},
\end{align}
where $\beta$ is the inverse temperature.
Following Refs. \onlinecite{Alt10, Sac15}, we decouple the interaction term on the top line of the action (\ref{appA:S}) with the Hubbard–Stratonovich transformation and introduce three non-local fields $G_\sigma(\tau,\tau')=-M^{-1} \sum_{i=1}^M \psi_{i\sigma}(\tau) \bar{\psi}_{i\sigma}(\tau')$, $G_c(\tau,\tau')=-N^{-1} \sum_{i=1}^N c_i(\tau) \bar{c}_i(\tau')$ together with $\Sigma_\sigma(\tau,\tau')$, $\Sigma_c(\tau,\tau')$ as the corresponding Lagrange multipliers:
\begin{align} \nonumber
S=&{}\int_0^\beta \!\!  d \tau \!\!  \int_0^\beta \!\!  d \tau' \Bigg[ \frac{M}{U}\delta(\tau-\tau')|\Delta|^2 - \!\sum_{i=1}^M\! \bar{\Psi}_i(\tau)\!
\begin{pmatrix} -\delta(\tau-\tau')\left(\partial_\tau -\mu\right)-\Sigma_\uparrow(\tau,\tau') & \delta(\tau-\tau')\Delta \\ \delta(\tau-\tau')\bar{\Delta} & -\delta(\tau-\tau')\left(\partial_\tau +\mu\right) -\Sigma_\downarrow(\tau,\tau') \end{pmatrix} \!\Psi_i(\tau') \\ \nonumber &{}-\sum_{i=1}^N \bar{c}_i(\tau) \big( -\delta(\tau-\tau')\partial_\tau -\Sigma_c(\tau,\tau')  \big) c_i(\tau')- M \!\sum_{\sigma=\uparrow,\downarrow}\!\!\left(\Sigma_\sigma(\tau,\tau') - \sqrt{\frac{N}{M}}\, t^2 G_c(\tau,\tau') \right)G_\sigma(\tau',\tau) \\  &{}-N \left(\Sigma_c(\tau,\tau')G_c(\tau',\tau)+\frac{J^2}{4}G_c(\tau,\tau')^4 \right)\!\Bigg] \label{appA:S_nlc},
\end{align}
where $\bar{\Psi}_i=\begin{pmatrix}\bar{\psi}_{\uparrow i} & \psi_{\downarrow i}\end{pmatrix}$ and $\Psi_i=\begin{pmatrix}\psi_{\uparrow i} & \bar{\psi}_{\downarrow i} \end{pmatrix}^T$ are Nambu spinors.
Integrating out fermions and assuming constant $\Delta$, we get:
\begin{align} \nonumber
S=&{}\frac{\beta M}{U}|\Delta|^2 -\!M\!\!\sum_{n=-\infty}^{+\infty}\!\!\! \log \bigg[\! \left(\mathrm{i}\omega_n-\Sigma_\uparrow(\mathrm{i}\omega_n)+\mu \right)\left(\mathrm{i}\omega_n-\Sigma_\downarrow(\mathrm{i}\omega_n) -\mu\right)-|\Delta|^2\bigg] -\!N\!\!\sum_{n=-\infty}^{+\infty}\!\!\!\log \bigg[\mathrm{i}\omega_n -\Sigma_c(\mathrm{i}\omega_n) \bigg] \\  &{}-\int_0^\beta \!\! d \tau \!\! \int_0^\beta \!\! d \tau' \Bigg[  M\!\!\sum_{\sigma=\uparrow,\downarrow}\!\!\left(\Sigma_\sigma(\tau,\tau') - \sqrt{\frac{N}{M}}\, t^2 G_c(\tau,\tau') \!\right)\!G_\sigma(\tau',\tau)\! + N \!\left(\Sigma_c(\tau,\tau')G_c(\tau',\tau)+\frac{J^2}{4}G_c(\tau,\tau')^4 \right)\! \Bigg] \label{appA:S_nlc_1},
\end{align}
where $\omega_n = \pi (2 n+1)/\beta$ are Matsubara frequencies. 
In the limit of $M$, $N \gg 1$, the saddle-point equations are:
\begin{align}
\Sigma_\uparrow(\tau)&{}= \frac{t^2}{\sqrt{p}} G_c(\tau), \;\;\; \Sigma_\downarrow(\tau)= \frac{t^2}{\sqrt{p}} G_c(\tau) \label{appA:Sud},\\
\Sigma_c(\tau)&{}=J^2 G_c(\tau)^3 +\sqrt{p} \, t^2 \!\!\sum_{\sigma=\uparrow,\downarrow}\!\! G_\sigma(\tau)\label{appA:Sc},\\
G_\uparrow(\mathrm{i}\omega_n)&{}=\frac{\mathrm{i}\omega_n -\mu-\Sigma_\downarrow(\mathrm{i}\omega_n)}{\left(\mathrm{i}\omega_n-\Sigma_\uparrow(\mathrm{i}\omega_n)+\mu \right)\left(\mathrm{i}\omega_n-\Sigma_\downarrow(\mathrm{i}\omega_n) -\mu\right)-|\Delta|^2}\label{appA:Gu},\\
G_\downarrow(\mathrm{i}\omega_n)&{}=\frac{\mathrm{i}\omega_n +\mu-\Sigma_\uparrow(\mathrm{i}\omega_n)}{\left(\mathrm{i}\omega_n-\Sigma_\uparrow(\mathrm{i}\omega_n)+\mu \right)\left(\mathrm{i}\omega_n-\Sigma_\downarrow(\mathrm{i}\omega_n)-\mu \right)-|\Delta|^2}\label{appA:Gd},
\end{align}
\begin{align}
G_c(\mathrm{i}\omega_n)^{-1}&{}=\mathrm{i}\omega_n -\Sigma_c(\mathrm{i}\omega_n)\label{appA:Gc},\\
\frac{1}{U}&{}=  \!\frac{1}{\beta}\!\sum_{n=-\infty}^{+\infty}\! \frac{1}{\left(\omega_n +\mathrm{i}\Sigma_\uparrow(\mathrm{i} \omega_n)-\mathrm{i}\mu \right)\left(\omega_n +\mathrm{i}\Sigma_\downarrow(\mathrm{i} \omega_n)+\mathrm{i} \mu \right) +|\Delta|^2}\label{appA:Delta},
\end{align}
where we introduced the parameter $p=M/N$ representing the amount of the SYK ``impurities'' in the superconductor sector.

We exclude the self-energies $\Sigma_{\sigma}$ (\ref{appA:Sud}), so that one obtains four Schwinger-Dyson equations:
\begin{align}
\Sigma_c(\tau)&{}=J^2 G_c(\tau)^3 +2 \sqrt{p} \, t^2  G_+(\tau)\label{appA:Sc1},\\
G_+(\mathrm{i}\omega_n)&{}=\frac{\mathrm{i}\omega_n -\frac{t^2}{\sqrt{p}} G_c(\mathrm{i}\omega_n)}{\left(\mathrm{i}\omega_n-\frac{t^2}{\sqrt{p}} G_c(\mathrm{i}\omega_n) \right)^2-\mu^2 -|\Delta|^2}\label{appA:Gud},\\
G_c(\mathrm{i}\omega_n)^{-1}&{}=\mathrm{i}\omega_n -\Sigma_c(\mathrm{i}\omega_n)\label{appA:Gc1},\\
\frac{1}{U}&{}=   \!\frac{1}{\beta}\!\sum_{n=-\infty}^{+\infty}\! \frac{1}{\left(\omega_n +\frac{\mathrm{i} t^2}{\sqrt{p}} G_c(\mathrm{i}\omega_n)\right)^2+\mu^2 +|\Delta|^2}\label{appA:Delta1},
\end{align} 
where the latter one (\ref{appA:Delta1}) is a modified BCS gap equation \cite{Alt10} and $G_+=\frac{1}{2}\left(G_\uparrow+G_\downarrow\right)$.

\section{Saddle-point numerical analysis}\label{appB}

\subsection{The algorithm}

To solve the equations (\ref{appA:Sc1}-\ref{appA:Delta1}), we use an iterative approach that is equivalent to finding the fixed point (the point to which the iterative procedure converges) of the operator $\hat{T}$ representing the Schwinger-Dyson equations (\ref{appA:Sc1}-\ref{appA:Gc1}) set on a fixed grid of Matsubara frequencies. \cite{comment_app} One starts with an empty seed $G^0$ and applies iterations
\begin{align}
G^{k+1} = \hat{T}G^{k} \label{eq:successive}
\end{align}
until
\begin{equation}
\n{G^{k+1} - G^{k}} \leq \varepsilon, \label{eq:criteria}
\end{equation}
where we set the precision to $\varepsilon = 10^{-4}$ and $\n{\cdot}$ denotes the euclidean norm of the vector. 

The straightforward approach (\ref{eq:successive}) converges rarely. One improves convergence modifying (\ref{eq:successive}) as
\begin{align}
G^{k+1} = \lambda G^{k} + (1-\lambda)\hat{T}G^{k}, \label{eq:weighted_successive}
\end{align}
where $0<\lambda<1$ is a tunable parameter. This particular approach (\ref{eq:weighted_successive}) has been used to compute the Green's function of the SYK model. \cite{Mal16_1} However, the convergence of the algorithm (\ref{eq:weighted_successive}) may sufficiently slow down when one considering extra Schwinger-Dyson equations coupled to those of the bare SYK model or expands the parameter space. In our case, that happens due to coupling of the SYK model to a superconductor. 
To cope with this problem, we suggest using the adaptive golden ratio algorithm, \cite{Mal19_alg} where the weight $\lambda$ is not fixed but automatically adjusted to the local properties of the operator $\hat{T}$:  
\begin{align}
\lambda_k &= \min\left\lbrace \frac{10}{9}\lambda_{k-1},\ \frac{9}{16
\lambda_{k-2}}\frac{\n{G^k-G^{k-1}}^2}{\n{G^k - \hat{T}G^k- G^{k-1}+\hat{T}G^{k-1}}^2}\right\rbrace, \label{lambda_k}\\
\bar{G}^k & = \frac{G^k+ 2 \bar{G}^{k-1}}{3}, \label{Gbar_k} \\
G^{k+1}& = \bar{G}^{k} - \lambda_k G^k +\lambda_k \hat{T}G^k. \label{G_k}
\end{align}
Above we introduce $\bar{G}$ as an auxiliary function that requires $\bar{G}^0 = G^1$ and $\lambda_0=\lambda_{-1}>0$.
Computationally, the algorithm (\ref{lambda_k}-\ref{G_k}) is of the same complexity as (\ref{eq:successive}) and (\ref{eq:weighted_successive}),
while the adaptive step allows for a significant speedup. 

We treat the pairing gap $\Delta$, the temperature $T$, and the coupling strength $t$ that enter the equations (\ref{appA:Sc1}-\ref{appA:Gc1}) as an external set of parameters. Once the Green's functions are found within the procedure (\ref{lambda_k}-\ref{G_k}), we choose the data that satisfies the self-consistency relation (\ref{appA:Delta1}) to produce the finite-temperature phase diagrams.

\subsection{Precision and grid}

Matsubara frequencies $\omega_n = \pi T(2 n+1)$ define a natural discrete grid. We set the ultraviolet cut-off $N$ such that $n\in \left[ -N, -N+1, \ldots, N-1, N\right]$, where the reliable $N$ is of the order $10^4$--$10^5$ with the accuracy criteria (\ref{eq:criteria}) $\varepsilon = 10^{-4}$. The numerical analysis becomes more demanding as one enters the low-temperature regime in the vicinity of the critical coupling.
We reach the lowest temperature of $T \sim 10^{-3}$  using $N = 1.5\times 10^6$, with a main computational bottleneck coming from the computer memory. Also, the computation of the lowest critical temperatures requires an increase of the accuracy for the self-consistency condition (\ref{appA:Delta1}) and $\varepsilon$ (\ref{eq:criteria}) to $10^{-5}$--$10^{-6}$.

\begin{figure*}[tb!!]
\center
\includegraphics[width=0.488\linewidth]{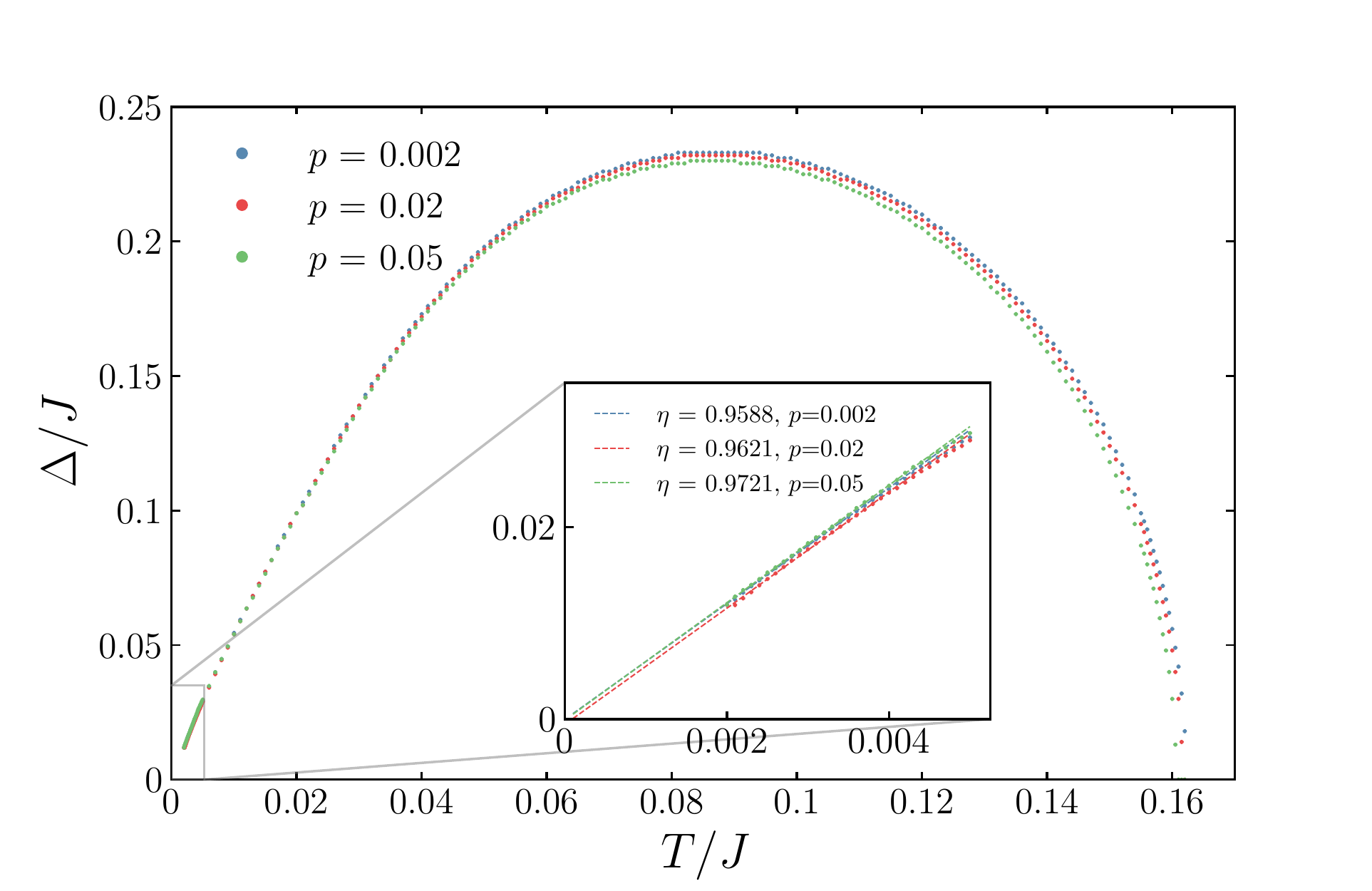} \quad
\includegraphics[width=0.488\linewidth]{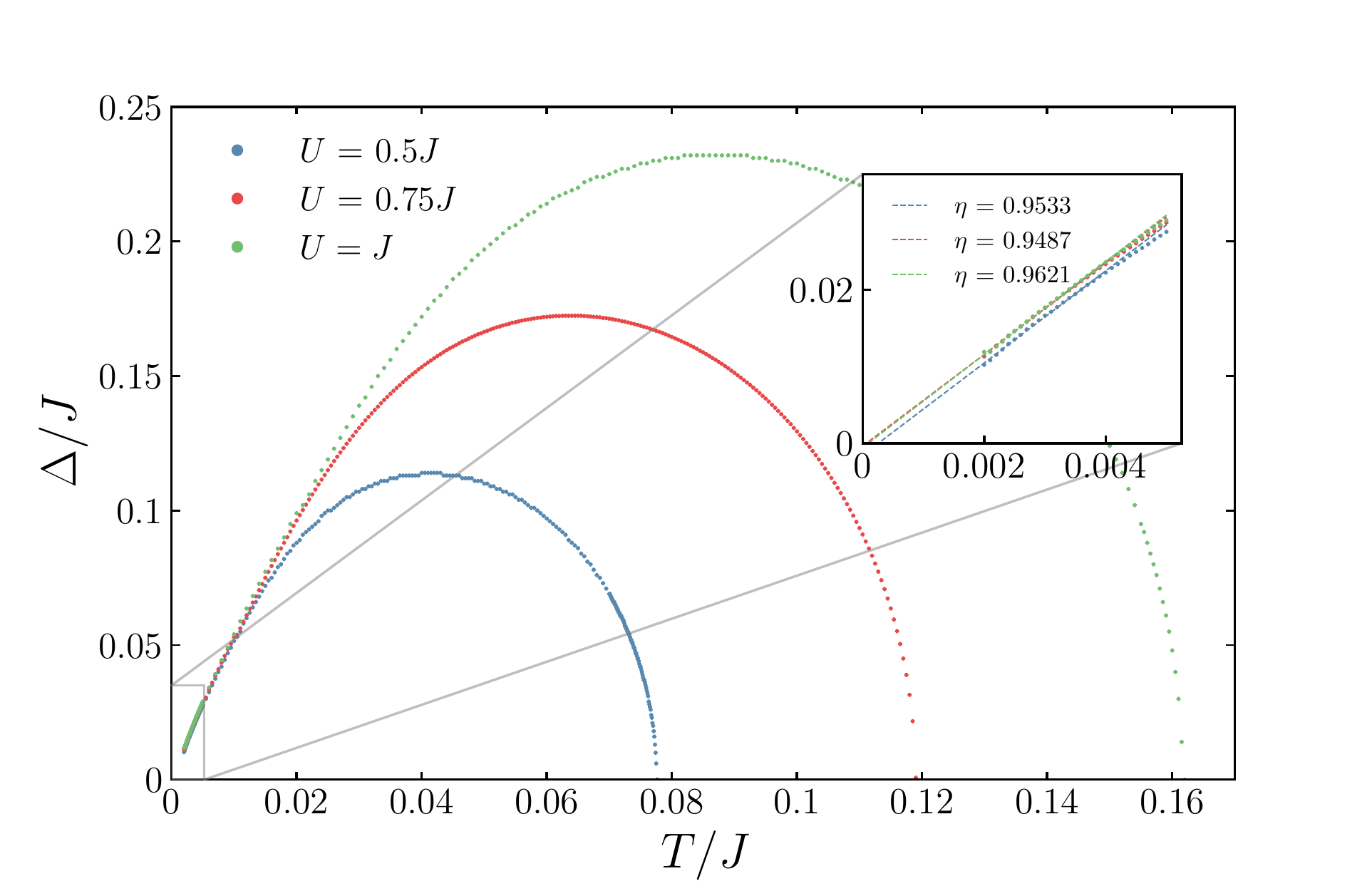}
\caption{\small \label{fig:delta_critical} \textbf{The pairing gap} as a function of temperature at the critical coupling. \textbf{Left panel}: fixed $U=J$. \textbf{Right panel}: fixed $p=0.02$.}
\end{figure*}

\begin{table}[tb!!]
  \begin{tabular}{ | c | c | c | c | c | c |}\hline
    $p$ & $0.002$ & $0.02$ & $0.02$ & $0.02$  & $0.05$  \\ \hline
    $U$ &     $1.0$    &    $1.0$       & $0.75$ & $0.5$   &  $1.0$     \\ \hline
    $t_c$ & $0.0710112 J$ & $0.126827 J$ & $ 0.10057J$ & $0.07294 J$ &  $0.1607 J$  \\ \hline
    $\eta$ & $0.9588$ & $0.9621$ &$0.9487$& $0.9533$ &  $0.9721$  \\ \hline
    $\delta$ &   $-2.96\times10^{-5}$   &     $-5.47\times10^{-4}$   & $-3.73\times 10^{-4}$ & $-1.54\times 10^{-3}$  &    $-9.86\times10^{-5}$   \\ \hline
  \end{tabular}
\caption{\small \label{tab:tp} The values of the \textbf{critical coupling} and the interpolation parameters for given $p$ and $U$.}
\end{table}

One of the objectives of this manuscript is to study the pairing gap at the critical coupling and low temperatures. In this regime, the gap grows linearly in temperature as shown in Figure \ref{fig:delta_critical}. 
The critical coupling $t_c$ is found as a condition when the off-set $\delta$ of the interpolating function $\Delta = 2\pi\eta\, T + \delta$ vanishes (see numerical values in Table \ref{tab:tp}). The system is sensitive to the coupling changes for small values of $p$, therefore, the precision of $t_c$ reaches $10^{-7}$ for $p=0.002$. 

\end{widetext}

\bibliography{refs}

\end{document}